\documentstyle[12pt]{article}
\def\hybrid{\topmargin -20pt    \oddsidemargin 0pt
        \headheight 0pt \headsep 0pt
        \textwidth 6.25in       
        \textheight 9.5in       
        \marginparwidth .875in
        \parskip 5pt plus 1pt   \jot = 1.5ex}

\hybrid

\def\baselinestretch{1.2}

\catcode`\@=11

\def\marginnote#1{}
\def\Jb{{\bar J}}
\def\a{\alpha}
\def\b{\beta}
\def\G{\Gamma}
\def\th{\theta}
\def\tht{\tilde{\theta}}
\def\d{\partial}
\def\db{\bar{\partial}}
\def\p{\phi}

\begin{document}

\newcommand{\inv}[1]{{#1}^{-1}} 

\renewcommand{\theequation}{\thesection.\arabic{equation}}
\newcommand{\beq}{\begin{equation}}
\newcommand{\eeq}[1]{\label{#1}\end{equation}}
\newcommand{\ber}{\begin{eqnarray}}
\newcommand{\eer}[1]{\label{#1}\end{eqnarray}}
\begin{center}
        September, 1993
			        \hfill    RI-153-93\\
                                \hfill    hep-th/9310016\\

\vskip .5in

{\large \bf Duality as a Gauge Symmetry and
Topology Change}\footnotemark \\

\footnotetext{Talk presented at the ``Strings 93'' conference,
May 24-29, 1993, Berkeley.
To appear in the Proceedings.}

\vskip .5in

{\bf Amit Giveon} \footnotemark \\

\footnotetext{e-mail address: giveon@vms.huji.ac.il}

\vskip .1in

{\em Racah Institute of Physics, The Hebrew University\\
  Jerusalem, 91904, ISRAEL} \\

\vskip .15in

\end{center}
\vskip .4in
\begin{center} {\bf ABSTRACT } \end{center}
\begin{quotation}\noindent
Duality groups as (spontaneously broken)
gauge symmetries for toroidal backgrounds, and their role in
($\infty$-dimensional) underlying string gauge algebras are reviewed.
For curved backgrounds, it is shown that there is a duality
in the moduli space of WZNW sigma-models,
that can be interpreted as a broken gauge
symmetry. In particular, this duality relates
the backgrounds corresponding to axially
gauged abelian cosets, $G/U(1)_a$, to vectorially gauged abelian cosets,
$G/U(1)_v$. Finally, topology change in the moduli space of WZNW
sigma-models is discussed.
\end{quotation}
\vfill
\eject
\def\baselinestretch{1.2}
\baselineskip 16 pt
\noindent
\section{Introduction}
\setcounter{equation}{0}

The new results in this talk are based on work with E. Kiritsis
\cite{GK}. I will discuss two issues in the moduli space of a WZNW
sigma-model:

\vskip .1in
\noindent
(a) There is a target space duality that can be interpreted, in the
effective action, as a broken gauge symmetry.
In particular, this duality relates the backgrounds corresponding to axially
gauged abelian cosets to vectorially gauged abelian cosets.

\vskip .1in
\noindent
(b) There is a topology change, namely, there are
deformations lines in the moduli space of sigma-models along
which the topology is changed.

\vskip .1in
The importance of the interpretation of duality as a gauge symmetry is:

\vskip .1in
\noindent
(1)  It shows that duality is an exact symmetry in string theory.
In particular, it shows that axial-vector duality is an exact symmetry;
this has  implications in the study of black-hole duality in string theory
\cite{bh},
and duality in cosmological string solutions \cite{cos,GPa}.

\vskip .1in
\noindent
(2) Dualities appear to be discrete symmetries (some kind of a ``Weyl
subgroup'') of a (infinite-dimensional) universal gauge algebra.

\vskip .1in
In section 2, I will explain point (2) reviewing known results on duality as
a gauge symmetry for toroidal
backgrounds. In  section 3, I will discuss the $J\Jb$ deformation of $SU(2)$
(or $``SL(2)"$) WZNW sigma-model, and a duality along this line will be
interpreted as a broken gauge symmetry. Finally, in section 4, I will discuss
topology change in the moduli space of WZNW sigma-models.

\section{Duality as a gauge symmetry and universal gauge algebras}
\setcounter{equation}{0}

\subsection{$R\to 1/R$ circle duality as a gauge symmetry}

I will first discuss the simplest case of a  single scalar field compactified
on a circle with radius $R$. The moduli space of circle compactifications is
the positive real line, describing all possible compactification radii $R>0$.
For each compactification radius, the conformal field theory (CFT)
has (at least) a $U(1)_L\times
U(1)_R$ affine symmetry, generated by the chiral and antichiral currents
$J$ and $\Jb$. The truly marginal deformation of the CFT, corresponding to
a change of the compactification radius from $R$ to $R+\delta R$,
is given by adding $\delta S=\delta(R^2)\int d^2z J\Jb$
to the worldsheet action.

A striking property of string theory is that there is a
``target-space duality'' symmetry
relating backgrounds with different geometries, that correspond to the same
CFT. In this case, the CFT corresponding to a radius $R$ circle is
equivalent to the CFT of a radius $1/R$ circle \cite{KY}.\footnote{
I choose $\alpha'=1$, where $\alpha'$ is the inverse string tension.
}

The conformal field theory at the self-dual point, $R=1$,
has an extended affine symmetry $SU(2)_L\times SU(2)_R$.
At this point there are three chiral currents $J^a$, $a=1,2,3$, and three
antichiral currents $\Jb^a$, generating the affine $SU(2)_L$ and $SU(2)_R$
symmetry, respectively. We shall refer to the point $R=1$ as the $SU(2)$
point. To deform the theory away from the $SU(2)$ point,
one may choose to identify $J\equiv J^3,\, \Jb\equiv \Jb^3$.
But this choice is not unique: any deformation of the type
\beq
(\sum_{a=1}^3\a_a J^a)(\sum_{b=1}^3\b_b \Jb^b)
\eeq{def}
is truly marginal.
The set of critical points, that can be reached from the $SU(2)$ point by
such conformal deformations, span a 5-dimensional surface in the
9-dimensional euclidean space generated by the couplings to $J^a\Jb^b$.
However, because different truly marginal deformations give rise to
the same CFTs, the dimension of the physical moduli space is 1.
This can be shown as follows:
At the $SU(2)$ point, $SU(2)_L$ and $SU(2)_R$ rotations of the
$J^a$'s and $\Jb^b$'s are symmetries of the CFT, and as a consequence, any
deformation of the type (\ref{def}) give rise to the same CFT given for an
appropriate $J^3\Jb^3$ deformation.

In particular, the duality transformation $R\to 1/R$ corresponds to the
Weyl transformation in $SU(2)_L$  that takes $J^3$ to $-J^3$ \cite{DHS}.
Infinitesimaly near the $SU(2)$ point, this corresponds to the
identification of the theory given by the deformation
$\delta\a J^3\Jb^3$ with
the theory given by the deformation $-\delta\a J^3\Jb^3$.

The discussion, so far, was from the worldsheet point of view. But in string
theory, worldsheet properties have a target-space analogue.
Indeed, coupling constants to
operators in the worldsheet action
become space-time fields in the effective action of string
theory. In particular, for a string compactified on a circle, we
expect to have a scalar field corresponding to the worldsheet
coupling to the $J\Jb$ deformation. The vacuum expectation values (VEVs) of
the scalar field correspond to the compactification radii.
At the particular VEV, corresponding to the $R=1$ point, there is
a gauge symmetry $SU(2)_L\times SU(2)_R$. At this point, extra scalar fields
and gauge fields become massless. Changing the VEV of scalar fields
away from this
point (or, equivalently, changing the compactification radius), the gauge
symmetry is spontaneously broken to $U(1)_L\times U(1)_R$, and, in
addition, there is a
symmetry corresponding to $R\to 1/R$ duality. In that sense, the
target space duality is a discrete symmetry of the spontaneously
broken $SU(2)$ gauge algebra. The $Z_2$ duality is the Weyl group of
$SU(2)$.

\subsection{$O(d,d,Z)$ dualities as gauge symmetries}

Compactifying the bosonic (heterotic) string on a $d$--dimensional torus,
the circle duality is generalized to a {\em duality group} isomorphic to
$O(d,d,Z)$ $(O(d,d+16,Z))$ \cite{GRV,SW,GMR}.

The duality group  may be interpreted (in the
effective action) as a residual discrete symmetry group of some
spontaneously broken ($\infty$--dimensional) universal gauge algebra.
Moreover, the $O(d,d,Z)$ dualities are expected to be some kind of a Weyl
subgroup of the underlying  algebra.
{}From the worldsheet point of view,
this can be shown as follows \cite{GMR2}: When $d>1$ there is an infinite
number of points in the moduli space of $d$--tori backgrounds that have
extended affine symmetries. For instance, when $d=2$
there is an infinite number of
points, in the moduli space, where the affine $U(1)^2$
symmetry is extended to $SU(2)\times SU(2)$ or to $SU(3)$.
Now, any product of Weyl reflections acting on conformal deformations
around points with extended symmetries relate geometrically different toroidal
backgrounds that correspond to the same CFT. It turns out that any such
``Weyl reflections'' product
is an element of $O(d,d,Z)$. Moreover, any element of
$O(d,d,Z)$ correspond to a product of Weyl reflections.\footnote{More
precisely, every element of $O(d,d,Z)$ is a product of Weyl reflections in
the moduli space of ($d+1$)--tori \cite{GMR2}.}

These worldsheet properties are realized in the target space in an
intriguing way.
In the effective action, one  expects to find a ($\infty$-dimensional)
gauge algebra, that is spontaneously broken for any VEV of the scalar
fields to an appropriate (finite-dimensional) gauge group, and residual
discrete symmetries generating the duality group. This is the
interpretation of target space dualities as the residual discrete
symmetries of a spontaneously broken gauge algebra.

Such an effective action was constructed in ref. \cite{GP}
for the $d=4,\, N=4$ heterotic string,
{\em i.e.}, the toroidal compactification of the heterotic string to four
dimensions.
The $\infty$--dimensional gauge algebra, called the ``Duality Invariant
String Gauge algebra'' (DISG) is isomorphic to
\beq
{\rm Duality} \, {\rm Invariant} \,{\rm String} \,{\rm Gauge} \,
{\rm algebra} \sim ``LatticeAlg"(\G^{6,22}),
\eeq{disg}
namely, the algebra of dimension 1 operators in the CFT of 28 chiral
scalars,
compactified on an even-self dual lorentzian lattice with signature (6,22).
The infinite-dimensional gauge symmetry is spontaneously broken for {\em
any} VEV of the scalar fields to a finite-dimensional gauge algebra
(typically $U(1)^6\times E_8\times E_8$),
and a group of residual discrete symmetries: the $O(6,22,Z)$ target-space
dualities.

This duality group is some kind of a Weyl subgroup of the underlying
gauge algebra. The subgroup of the the duality group, that fixes a point in
the moduli space of toroidal backgrounds, is related to the
Weyl group of the enhanced (finite-dimensional) gauge
symmetry at the fixed point. To get points with
large enhanced gauge symmetries and
large duality subgroups that fix such points, it is useful to compactify
time as well.

\subsection{Compactifying time}

By compactifying
all the coordinates -- including timelike dimensions -- on a torus,
one recovers points in the moduli space of backgrounds with large gauge
symmetries; these correspond, in the worldsheet,
to large on--shell algebras of dimension (1,0)
or (0,1) operators. Some  backgrounds are fixed points of large
duality subgroups. Compactifying time is, therefore,
useful in the search for an underlying universal
gauge algebra of string theory.

Such a program was initiated for the critical $N=2$ string, compactified
completely on a torus $T^{2,2}$ \cite{GS}.
A distinguished point in the moduli space of (2,2) toroidal backgrounds is
the one where the Narain lattice $\G^{4,4}$ is the direct
sum of a $(2,2)_L$ even self-dual lorentzian lattice of left-movers, and
a $(2,2)_R$ even self-dual lorentzian lattice of right-movers
$$
{\rm Maximal}\, {\rm Extended} \, {\rm Symmetry} \, {\rm Point}:\,\,\,
\G^{4,4}_{MES}=\G_L^{2,2}\oplus \G_R^{2,2}.
$$
At this point the extended on--shell gauge algebra is infinite-dimensional:
it is generated by the area-preserving diffeomorphisms of null 2-tori in
the $(2,2)_L$ (and $(2,2)_R$) torus.

A universal gauge algebra is a {\em background independent} algebra
that is a minimal Lie algebraic closure which  contains {\em all} the
on--shell algebras.
Candidates for  universal gauge algebras of the $N=2$ string
were presented in ref. \cite{GS}. These are:

\vskip .1in
\noindent
(1)  ``Lattice-Algebra"$(\G^{4,4})$.

\vskip .1in
\noindent
(2) Volume-Preserving-Diffeomorphisms$(T^{4,4})$.

\vskip .1in
\noindent
Here $T^{4,4}\equiv R^{4,4}/\G^{4,4}$
is the Narain torus defined by the Narain lattice
$\G^{4,4}$. The first candidate is the analogue of the
DISG for the $N=2$ string; the duality group $O(4,4,Z)$ is some kind of its
Weyl subgroup. The second candidate seems more natural for the $N=2$ string.

In a recent paper \cite{M}, Moore has presented a candidate for a universal
symmetry algebra of the bosonic string. One starts by
compactifying all coordinates on a (1,25) torus. A distinguished point in
the moduli space of toroidal backgrounds is the one where the Narain
lattice   $\G^{26,26}$ is the direct
sum of a $(25,1)_L$ even self-dual lorentzian lattice of left-movers, and
a $(1,25)_R$ even self-dual lorentzian lattice of right-movers
$$
{\rm Maximal}\, {\rm Extended} \, {\rm Symmetry} \, {\rm Point}:\,\,\,
\G^{26,26}_{MES}=\G_L^{25,1}\oplus \G_R^{1,25}.
$$
At this point the extended on--shell gauge algebra is infinite-dimensional:
it is the direct sum of the
``Fake Monster Lie Algebra" ($\sim$ ``Lattice-Algebra"$(\G^{1,25})$)
of left-movers and right-movers.

The candidate for a universal gauge symmetry is a modification of the DISG
algebra (on a modified Narain lattice, needed to include properly
the left-moving enhanced symmetries together with the right-moving enhanced
symmetries):
Universal Symmetry $\sim \tilde{DISG}\equiv
``LatticeAlg"(\tilde{\G}^{26,26})$.
For more details see ref. \cite{M}.

\section{$J\Jb$ deformation of $SU(2)$
(or $SL(2)$) WZNW sigma-model, and a duality as a broken gauge symmetry}
\setcounter{equation}{0}

The discussion in section 2 is restricted to toroidal backgrounds.
What about target space dualities in the moduli space of {\em curved}
backgrounds? Can they  be interpreted as
some (spontaneously broken) gauge symmetries?
Here I shall discuss, from the worldsheet point of view,
a particular element of the duality group in the
moduli space of WZNW models. For simplicity, I will discuss the simplest
non-trivial case, namely, duality acting on the conformal deformations line
of $SU(2)$ (or $``SL(2)"$) WZNW models. In this note I will only present the
results; the details appear in \cite{GK}.

The action for $SU(2)_k$ (in a particular parametrization of the group
elements) is given by
\ber
S[x,\th,\tht]=\frac{k}{2\pi}\int d^2 z
(\d\th, \d\tht, \d x)\left( \begin{array}{clcr}
                                          \sin^2x & \cos^2x & 0\\
                                         -\cos^2x & \cos^2x & 0\\
                                          0 & 0 & 1
                                         \end{array}\right)
\left(\begin{array}{clcr} \db \th\\ \db\tht \\ \db x\end{array} \right)
\nonumber \\
-\frac{1}{8\pi}\int d^2 z \p_0 R^{(2)},
\eer{Ssu2}
where $x\in [0,\pi/2)$ and $\th,\tht\in [0,2\pi)$. We will refer to the
matrix in (\ref{Ssu2}) as ``the background matrix'' $E$.

The action $S$ (\ref{Ssu2}) is manifestly invariant under the
$U(1)_L\times U(1)_R$ affine symmetry generated by the currents
\beq
J=k(-\sin^2x\d\th +\cos^2x\d\tht)\qquad
\Jb=k(\sin^2x\db\th +\cos^2x\db\tht)\; .
\eeq{JJ}

It is possible to deform the action $S$ to new conformal backgrounds by
adding to it any marginal  deformation. We will focus on the marginal
deformation $J\Jb$:
\beq
S\to S+\a\int d^2z J\Jb.
\eeq{adef}
This deformation is equivalent to a particular 1--parameter sub-family
of $O(2,2,R)$ rotations acting on the background matrix and dilaton in
(\ref{Ssu2}). Parametrizing this family by $0\leq R < \infty, $ one finds
the line of exact CFTs with background matrices:
\beq
kE_R=k\left( \begin{array}{clcr}
                          \tan^2x/\Delta & 1/\Delta & 0\\
                          -1/\Delta & R^2/\Delta & 0\\
                                          0 & 0 & 1
                                         \end{array}\right),
\qquad \Delta=1+R^2\tan^2x.
\eeq{ER}
The dilaton also transforms under the deformation (see \cite{GK}).

Some special points along the $R$-line of deformations are:

\vskip .1in
\noindent
(1) At $R=1$, one recovers the original $SU(2)$ point (\ref{Ssu2}).

\vskip .1in
\noindent
(2) At $R\to\infty$, the background corresponds to a direct product of an
axially gauged $SU(2)/U(1)_a$ sigma-model \cite{BCR},
and a non-compact free scalar field.

\vskip .1in
\noindent
(3) At $R\to 0$, the background corresponds to a direct product of a
vectorially gauged $SU(2)/U(1)_v$ sigma-model,
and a non-compact free scalar field.

\vskip .1in

We now arrive to the important point of this part. The Weyl reflection
$J\to -J$ is given by a continuous group rotation at the WZNW point. This
implies that $J\to -J$ is  a symmetry at the WZNW point, and therefore, the
infinitesimal deformation $S_{WZNW}\to S_{WZNW}+\delta\a\int d^2z J\Jb$
is the same CFT as the one given by the
deformation $-\delta\a\int d^2z J\Jb$.

The points $\delta\a$ and $-\delta\a$ along the $\a$--modulus (\ref{adef})
are  {\em the same CFT}.
In string theory we say: $\delta\a$ and
$-\delta\a$ are related by a residual $Z_2$ symmetry of the broken gauge
symmetry (of the enhanced symmetry point); the residual symmetry is the
target space duality.
This symmetry can be integrated to finite $\a$, giving rise to a $Z_2\in
O(2,2,Z)$ duality\footnote{
The $O(d,d,Z)$ group is a duality (sub-)group also for {\em curved}
backgrounds with $d$ abelian
isometries \cite{GR}.}
along the $R$--line (\ref{ER}).
The action of duality on the background $E_R$ is:
\beq
{\rm Duality}\, {\rm along}\, {\rm the}\, R{\rm -line}:\,\, E_R\to E_{1/R}
\qquad i.e. \qquad R\to 1/R.
\eeq{Edual}

In particular, the boundary points $R\to 0$ and $R\to \infty$ are the same
CFT. As a consequence, axial-vector duality of $SU(2)/U(1)$ and
$SL(2)/U(1)$ is exact, and corresponds to a residual discrete symmetry of
the broken gauge algebra.

I shall end this part with few remarks:

\vskip .1in
\noindent
(a) In the $SU(2)$ ($SL(2)$) case, the CFT along the $R$-line is a
$Z_k$ orbifoldization of a compact (non-compact) parafermionic theory
and a free scalar field with radius $\sqrt{k}R$.

\vskip .1in
\noindent
(b) In the $SU(2)$ case, the modular invariant $R$-dependent
genus one partition function can be written.

\vskip .1in
\noindent
(c) The results of this section can be extended to {\em any} group $G$
\cite{GK}.

\section{Topology change in the space of WZNW sigma-models}
\setcounter{equation}{0}

Along the $0<R<\infty$ deformations line of the $SU(2)$--WZNW sigma-model,
the topology of the background space is of the three sphere.
However, at the boundaries ($R=0,\infty$),
the topology is changed to that of
a product of a two-disc (corresponding to the
$SU(2)/U(1)$ conformal background) with a circle whose
radius shrinks to 0. In the $``SL(2)"$ case, the background at the boundary
is the direct product of
a degenerated circle with a semi-infinite cigar (at
$R\to\infty$), or the infinite trumpet (at $R\to 0$); these correspond to
the dual pair of the 2-$d$ euclidean black-hole backgrounds.

A topology change at the boundary of moduli space is not surprizing.
However, the $R$-line of deformations is not the full story in the moduli
space of $G$-WZNW sigma-models.
In fact, any conformal sigma-model with $d$ abelian isometries can be
transformed  to a new conformal background by
$O(d,d,R)$ rotations \cite{V,GR}.\footnote{In the bosonic string, the
$O(d,d)$ rotations give only the leading order in $\a'$ of the conformal
backgrounds; but, there are higher order corrections that make them exact
\cite{GR,GPa,K}.}
In this bigger moduli space of sigma-models, there
are some more interesting deformation lines, along which the topology might
change.

For example, in the moduli space of the
$SU(2)$--WZNW sigma-model, there is a
one-parameter sub-family of $O(2,2,R)$ rotations that generate backgrounds
with metric \cite{GK}
\ber
ds^2(\a)=\frac{k}{\Delta(\a)}[\sin^2x d\th^2+\cos^2x d\tht^2]+kdx^2,
\nonumber\\
\Delta=\cos^2 \a \cos^2x+(\cos\a+k\sin\a)^2\sin^2x .
\eer{line}
(There are also non-trivial dilaton and torsion along this $\a$--line).
At the point $\a=0$ the background includes a metric of
the $SU(2)_k$ group manifold $S^3$ (as well as an antisymmetric background).
Along the line $0<\a <\pi/2$ the background includes the
metric (\ref{line}) with the topology
of $S^3$ (as well as an antisymmetric background and a dilaton field).
At the point $\a=\pi/2$ the background metric is
\beq
ds^2(\a =\pi/2)=\frac{1}{k}d\th^2+\frac{1}{k}\cot^2xd\tht^2+kdx^2.
\eeq{aline}
At this point the manifold has a topology of $D_2\times S^1_{1/k}$,
where $D_2$
is a two-disc and $S^1_{1/k}$ is a circle with radius $R^2=1/k$.
One may continue to deform this theory by, for example, changing the
compactification radius $R$ of the free scalar field $\th$.

It is remarkable that (for integer $k$) the neighborhood of the point
$\a=\pi/2$ is mapped to the neighborhood of
the point $\a=0$
by an element of $O(2,2,Z)$, namely, a target space generalized duality.
Therefore, a region in the moduli space, where a topology change occurs,
is mapped to a region where there is no topology change at all. A similar
phenomenon happens for more complicated examples in the moduli space of
Calabi-Yau compactifications \cite{CY}.

Finally, let me make few comments:

\vskip .1in
\noindent
(a) The sigma-models along the $\a$--line (\ref{line})
have  conical singularities.
Therefore, to make sense of the CFTs along the $\a$--line one should
understand CFTs corresponding to backgrounds with (non-orbifold)
conical singularities.

\vskip .1in
\noindent
(b) After the topology is changed at (\ref{aline}), by
deforming the compactification radius of the free scalar field $\th$,
one does not get rid of the curvature singularity encountered at the point
where the topology is changed.

\vskip .1in
\noindent
(c) To cure both problems, one should look at the moduli space of WZNW
sigma-models in higher dimensions; for instance, the sigma-model moduli
space discussed for cosmological backgrounds in \cite{GPa}.

\vskip .3in \noindent
{\bf Acknowledgements} \vskip .2in \noindent
I thank S. Elitzur for his remarks on the manuscript.
This work is supported in part by the BSF - American-Israeli Bi-National
Science Foundation.

\end{document}